\documentclass[12pt]{article}
\usepackage[cp1250]{inputenc}
\usepackage[verbose,a4paper,tmargin=0.5in,lmargin=1in,rmargin=1in]{geometry}
\usepackage[pdftex]{graphicx}
\usepackage{amsfonts}
\usepackage{indentfirst}
\usepackage{latexsym}
\usepackage{amsmath}
\usepackage{amsthm}
\usepackage{amssymb}
\usepackage{psfrag}
\usepackage{eufrak}
\usepackage[mathscr]{eucal} 
\usepackage{color}
\usepackage{cite}
\usepackage{verbatim}

\title{On twisting type $[\textrm{N}] \otimes [\textrm{N}]$ Ricci flat complex spacetimes with two homothetic symmetries}

\author{$\textrm{Adam Chudecki}^{*}$, $\textrm{Maciej Przanowski}^{**}$}

\begin{document}

\maketitle

$*$ Center of Mathematics and Physics, Lodz University of Technology, 
\newline
$\ \ \ \ \ $ Al. Politechniki 11, 90-924 Łódź, Poland, adam.chudecki@p.lodz.pl

$**$ Institute of Physics, Lodz University of Technology, 
\newline
$\ \ \ \ \ $ Wólczańska 219, 90-924  Lodz, Poland, maciej.przanowski@p.lodz.pl
\newline
\newline
\newline
\textbf{Abstract}. 
\newline
$\mathcal{HH}$ spaces of type $[\textrm{N}] \otimes [\textrm{N}]$ with twisting congruence of null geodesics defined by the 4-fold undotted and dotted Penrose spinors are investigated. It is assumed that these spaces admit two homothetic symmetries. The general form of the homothetic vector fields are found. New coordinates are introduced which enable us to reduce the $\mathcal{HH}$ system of PDEs to one ODE on one holomorphic function. In a special case this is a second-order ODE and its general solution is explicitly given. In the generic case one gets rather involved fifth-order ODE.
\newline
\newline
PACS numbers: 04.20.Cv, 04.20.Jb, 02.40.Ky


\setcounter{equation}{0}
\section{Introduction}

There is an anecdote about Werner Heisenberg that once he said "When I meet God, I am going to ask him two questions: Why relativity? And why turbulence? I really believe he will have an answer for the first". We are quite sure, that many of relativists will specify the first question and they are going to ask "Why twisting type N?".

From the theory of gravitational radiation and, in particular, from the \textsl{Sachs peeling theorem} \cite{1,2,3,4,5,6,7,8,9,10} we conclude that far away from a bounded source of gravitational radiation the gravitational field is approximately of the Petrov - Penrose type N. So, the natural question is to find a vacuum type N solution of Einstein equations which could represent gravitational radiation from some bounded source. At the first glance it seems that we are in a very good position since a lot of vacuum type N solutions are known. In fact, all vacuum type N metrics admitting the \textsl{hypersurface-orthogonal congruence of shearfree null geodesics ($\equiv$ nontwisting congruence of rays}) are known \cite{9}. Unfortunately, all type N vacuum metrics admitting nontwisting but expanding congruence of rays i.e. the Robinson - Trautman metrics \cite{9,10,11,12,13} which eventually could correspond to some gravitational radiation field from a bounded source are singular along some line(s) and, consequently, they cannot describe none of such gravitational radiation. The only reasonable way out of this difficulty is to consider the vacuum type N solution with \textsl{twisting congruence of rays} ($\equiv V_{[a;b]}V^{[a;b]} \ne 0$ everywhere; with $V^{a}$ being the vector field tangent to the congruence, the semicolon ";" denotes the covariant derivative and the bracket $[ \cdot, \cdot]$ stands for the antisymmetrization). However, this turned out to be a hard problem and until now we know only one twisting type N solution of vacuum Einstein equations, the \textsl{Hauser solution}, found by Isidore Hauser in 1974 \cite{9,14,15}. The Hauser solution also cannot represent the radiation field from bounded source since it is not asymptotically flat and is singular on some hypersurfaces. A lot of effort was made to find other twisting vacuum type N metrics but without the ultimate success \cite{9,12,13,16,17,18,19,20,21,22,23,24,25,26,27,28,29,30,31,32,33,34,35,36,37,38,39,40,41,42}. Since the Hauser metric admits two homothetic vector fields \cite{13,16,17} several authors attack the twisting vacuum type N problem under the additional assumption that the space considered admits two homothetic symmetries \cite{13,16,18,21,22,23,28,30,35,36,38}. Although such an approach simplifies considerably the field equations which are now reduced to one ODE or to the system of ODEs, however, the final equation(s) is (are) very much complicated. 

Another approach is based on approximative methods \cite{26,29,31,32,34,40}. It has been inspired by the Robinson - Trautman solution \cite{9,11} and by the Sommers and Walker work \cite{43} on linearized Hauser equation, but it did not give any conclusive answer concerning the properties (the existence of singularities) of the twisting type N vacuum fields. 

Then, some authors suppose that a very much promising method for investigation the twisting type N problem follows directly from complex relativity \cite{20,24,27,33,38,42}. Here, the reasoning is in short so: Complex Einstein field equations for the \textsl{one-sided degenerate Ricci flat complex spacetime ($\equiv$ hyperheavenly or $\mathcal{HH}$ space)} can be reduced to one second order nonlinear PDE for one holomorphic function $W$ called the \textsl{key function} and defined in the $\mathcal{HH}$ space \cite{44,45,46}. Then one should specialize this PDE to the case when the $\mathcal{HH}$ space is of the \textsl{type N on both sides} i.e. both self-dual and anti-self-dual parts of the Weyl tensor are of the type N. Moreover, since we are interested in the twisting case we must assume that the congruence of complex null geodesic lines (rays) defined by the tangent vectors $a_{A}a_{\dot{A}}$, $A=1,2; \, \dot{A} = \dot{1}, \dot{2}$ (where $a_{A}$ and $a_{\dot{A}}$ are the 4-fold undotted and dotted, respectively, Penrose spinors) is twisting. Taking appropriate real slice \cite{47}, if it does exist, one gets a vacuum type $[\textrm{N}] \otimes [\textrm{N}]$ real metric of signature $(++--)$ or a vacuum type N real metric of Lorentzian signature admitting twisting congruence of rays. Obviously, we are mainly interested in that latter case as it describes a real gravitational field but also the first possibility is of some interest. The evident advantage of the "complex" approach to twisting type N problem is the fact that from the very beginning we deal with only one PDE on one function $W$. The success of this approach depends very much on appropriate choice of coordinates. A remarkably useful choice was proposed in a distinguished paper by Plebański and Torres del Castillo \cite{48}. Then the works \cite{20,24,27,38,42} inspired by the results of \cite{48} pushed the "complex" approach to the twisting type N a bit forward but still we are far from final solution. The present paper is devoted to further investigation of this problem. 

In section \ref{basic_formalism} we remind basic formalism of hyperheavenly space ($\mathcal{HH}$ space) theory. We mainly focus on the type $[\textrm{N}] \otimes [\textrm{N}]$ with expanding congruence of self-dual (SD) null strings. Definition of twisting type $[\textrm{N}] \otimes [\textrm{N}]$ $\mathcal{HH}$ space is given and the $\mathcal{HH}$ equation is explicitly written down. In section \ref{section_new_coordinates} new coordinates are introduced and this simplifies considerably the form of key function $W$ defining the metric of $\mathcal{HH}$ space. Then the $\mathcal{HH}$ equation in our new coordinates is brought to some system of PDEs, the \textsl{$\mathcal{HH}$ system of equations} (\ref{uklad_rownan_na_typN_bez_symetrii}). The relation between these coordinates and the ones employed in \cite{20,24,42,48} is analyzed. In section \ref{section_Homothetic_symmetries} we investigate the case when twisting type $[\textrm{N}] \otimes [\textrm{N}]$ $\mathcal{HH}$ space admits two homothetic vector fields. First one notes that without any loss of generality we can assume that one (at least) of that vectors is a Killing vector. Then we are able to solve so called \textsl{master equation} (\ref{master_equation}) and to find the explicit form of this Killing vector and of the additional homothetic vector. We show also that similarly as in real Lorentzian case \cite{13,16,17,18} also in the complex case the maximal dimension of the homothetic group is \textsl{two} and this group must be non-Abelian. One finds that the further investigation of the space with two homothetic symmetries depends very much on the properties of a crucial function $f$ (see (\ref{rozwiazanie_na_h})) which determines our new coordinate $h$. So one must consider two cases: first, when $\dfrac{\partial f}{\partial h}=0$ and the second, the \textsl{generic case}, when $\dfrac{\partial f}{\partial h} \ne 0$ (see (\ref{warunek_zerowosci_dotf}), (\ref{warunek_zerowosci_f}) and (\ref{warunek_niezerowosci_dotf})).

As is shown in section \ref{section_f_zero} the first case can be solved completely. We are able to reduce the $\mathcal{HH}$ system of equations to one second order linear ODE for one function $u=u(h)$. This equation can be brought to the hypergeometric equation. Finally, one gets the general solution for $u(h)$ as a power series (\ref{rozwiazanie_w_postaci_szeregu}) and then the metric can be also easily find as (\ref{metryka_przyklad_1}). The metric has a natural real slice but, unfortunately, this slice is of signature $(++--)$. It must be noted that all possible real slices of that metric are of signature $(++--)$. 

The generic case studied in section \ref{sekcja_f_niezero} is much more complicated. Here we were able to reduce the $\mathcal{HH}$ system of equations to one fifth order nonlinear ODE for one function, Eq. (\ref{HORROR_EQUATION}). This equation is extremely involved and until now we cannot find any solution. Therefore further effort is needed. Nevertheless, the obvious advantage of our approach is the reduction of $\mathcal{HH}$ system to a single ODE.


\setcounter{equation}{0}
\section{Basic formalism}
\label{basic_formalism}

In the paper we extensively use the $\mathcal{HH}$ (hyperheavenly) space formalism introduced by J.F. Plebański, I. Robinson, J.D. Finley and Ch. Boyer in the seminal works \cite{44,45,46}.

We define \textsl{$\mathcal{HH}$-space (hyperheavenly space)} as a 4-dimensional complex analytic differential manifold endowed with a Ricci flat holomorphic metric $ds^{2}$ such that the self-dual or anti-self-dual part of the Weyl tensor is algebraically degenerate. Of course, since the Ricci tensor vanishes the Weyl tensor is equal to the curvature tensor. From the complex Goldberg - Sachs theorem \cite{46,49,50,51} it follows that for each point $p$ of any $\mathcal{HH}$ space there exist an open neighborhood $U$ of $p$ and a 2-dimensional complex totally null self-dual or anti-self-dual, respectively, distribution $\mathcal{D}_{2}$ on $U$ which appears to be completely integrable and the integral manifolds of $\mathcal{D}_{2}$ constitute the congruence of totally null, totally geodesic, self-dual or anti-self-dual, complex 2-dimensional surfaces called \textsl{self-dual (SD) null strings} or \textsl{anti-self-dual (ASD) null strings}.

Since the considerations of the present paper are purely local we will identify the neighborhood $U$ with all the $\mathcal{HH}$ space. As has been already mentioned we are going to investigate some $\mathcal{HH}$ spaces of the type $[\textrm{N}] \otimes [\textrm{N}]$ what means that both SD and ASD parts of the Weyl tensor are of the type [N] and, by the complex Goldberg - Sachs theorem this is equivalent to the statement that there exist exactly one congruence of SD null strings and exactly one congruence of ASD null strings. Moreover, these congruences are uniquely defined by the "undotted" or "dotted" fourfold Penrose spinor, respectively.

As was observed in \cite{44,45,46} and then it has been studied in detail \cite{51,52}, the crucial geometrical characteristic of a congruence of null strings which very much influences the reduction of Einstein equations and the form off metric is whether the congruence is \textsl{expanding} or \textsl{nonexpanding}. We say that the congruence of null strings is \textsl{expanding} if the 2-dimensional complex distribution $\mathcal{D}_{2}$ defining this congruence is, at every point, not parallely propagated. If $\mathcal{D}_{2}$ is parallely propagated we say that the respective congruence of null strings is \textsl{nonexpanding}.

From reasons which will be explained soon (see the paragraph after Eq. (\ref{twist})) in this work we focus on the $\mathcal{HH}$ space of the type  $[\textrm{N}] \otimes [\textrm{N}]$ with expanding congruence of SD null strings.

The metric $ds^{2}$ of such a space can be written in the form \cite{45,46}
\begin{equation}
\label{metryka_forma_1}
ds^{2} = - \frac{1}{2} g_{A\dot{B}} \underset{s}{\otimes} g^{A \dot{B}}
\end{equation}
where the null tetrad $g^{A\dot{B}}$ reads
\begin{eqnarray}
\label{PRF_tetrad}
g^{2 \dot{A}} &=& - \sqrt{2} \phi^{-2} dq^{\dot{A}}
\\ \nonumber
g^{1 \dot{A}} &=& - \sqrt{2} (dp^{\dot{A}} - Q^{\dot{A}\dot{B}} dq_{\dot{B}} )
\end{eqnarray}
and the spinorial indices $A,B, \textrm{etc}.=1,2$; $\dot{A},\dot{B}, \textrm{etc}.= \dot{1},\dot{2}$; $(p^{\dot{A}}, q^{\dot{B}})$ are the coordinates and $Q^{\dot{A}\dot{B}}$ are the structural functions. [The spinorial indices are to be manipulated with the use of anti-symmetric 2-spinors
\begin{equation}
\nonumber
( \in_{AB} ) := \left[ \begin{array}{cc}
                            0 & 1   \\
                           -1 & 0  
                            \end{array} \right] =: ( \in^{AB} )
\ \ \ , \ \ \ 
( \in_{\dot{A}\dot{B}} ) := \left[ \begin{array}{cc}
                            0 & 1   \\
                           -1 & 0  
                            \end{array} \right] =: ( \in^{\dot{A}\dot{B}} )
\end{equation}
as follows
\begin{equation}
\nonumber
\Psi_{A} = \in_{AB} \Psi^{B} , \ \ \ \Psi_{\dot{A}} = \in_{\dot{A}\dot{B}} \Psi^{\dot{B}}, \ \ \
\Psi^{A} = \in^{BA} \Psi_{B} , \ \ \ \Psi^{\dot{A}} = \in^{\dot{B}\dot{A}} \Psi_{\dot{B}} \ ]
\end{equation}

Assume that the $\mathcal{HH}$ space is of the type $[\textrm{N}] \otimes [\textrm{any}]$ with expanding congruence of SD null strings. From the vacuum Einstein equations without a cosmological constant ($\equiv$ the Ricci flatness) one infers that $\phi$ can be put as $\phi = J_{\dot{A}} p^{\dot{A}}$, where $J_{\dot{A}}$ is a nonzero constant spinor and the structural functions $Q^{\dot{A}\dot{B}} = Q^{(\dot{A}\dot{B})}$ take the form 
\begin{equation}
\label{postac_QAB}
Q^{\dot{A}\dot{B}} = - [ \phi^{4} (\phi^{-3} W)_{p_{( \dot{A}}} ]_{p_{\dot{B})}}
\end{equation}
where the bracket $(\cdot, \cdot)$ stands for the symmetrization. The function $W=W(p^{\dot{A}}, q^{\dot{B}})$ is called the \textsl{key function}, and this is the basic object of the $\mathcal{HH}$ formalism as the vacuum Einstein equations reduce to one second-order nonlinear PDE on the function $W$. This equation is called the \textsl{expanding $\mathcal{HH}$ equation for the type $[\textrm{N}] \otimes [\textrm{any}]$}
\begin{equation}
\label{expanding_HH_equation}
\frac{1}{2} \phi^{4} (\phi^{-2} W_{p_{\dot{B}}})_{p_{\dot{A}}} (\phi^{-2} W_{p^{\dot{B}}})_{p^{\dot{A}}}  + \phi^{-1} W_{p_{\dot{A}} q^{\dot{A}}} = \frac{1}{2} \varkappa \phi + \gamma
\end{equation}
where $\varkappa = \varkappa (q^{\dot{A}})$, $\gamma = \gamma (q^{\dot{A}})$ are some functions of $q^{\dot{A}}$ only and, as in (\ref{postac_QAB}) and in the rest part of the paper $W_{p^{\dot{B}}} \equiv \dfrac{\partial W}{\partial p^{\dot{B}}}$, $W_{p^{\dot{B}} p^{\dot{A}}} \equiv \dfrac{\partial^{2} W}{\partial p^{\dot{B}} \partial p^{\dot{A}}}$,  $W_{p_{\dot{A}} q^{\dot{A}}} \equiv \dfrac{\partial^{2} W}{\partial p_{\dot{A}} \partial q^{\dot{A}}}$, ..., etc.

Spinorial image $C_{ABCD}$ of the SD part of the Weyl tensor reads
\begin{equation}
\label{SD_conformal_curvature}
C_{1111} = C_{1112} = C_{1122} = C_{1222} = 0, C_{2222} = \phi^{7} J^{\dot{B}} \gamma_{q^{\dot{B}}}
\end{equation}
and the spinorial image $C_{\dot{A}\dot{B}\dot{C}\dot{D}}$ of the ASD part of the Weyl tensor is given by 
\begin{equation}
\label{ASD_conformal_curvature}
C_{\dot{A}\dot{B}\dot{C}\dot{D}} = \phi^{3} \, W_{p^{\dot{A}}p^{\dot{B}}p^{\dot{C}}p^{\dot{D}}}
\end{equation}
From (\ref{SD_conformal_curvature}) and (\ref{ASD_conformal_curvature}) one quickly concludes that the $\mathcal{HH}$ space considered is of the type $[\textrm{N}] \otimes [\textrm{N}]$ iff
\begin{equation}
\label{warunki_na_niezerowa_krzywizne}
C_{2222} \ne 0 \ \Longleftrightarrow \ J^{\dot{B}} \gamma_{q^{\dot{B}}} \ne 0
\end{equation}
and there exists a nonzero spinor $a_{\dot{A}}$ such that
\begin{equation}
\label{rozklad_ASD_krzywizny_na_spinory}
W_{p^{\dot{A}}p^{\dot{B}}p^{\dot{C}}p^{\dot{D}}} = \Psi a_{\dot{A}} a_{\dot{B}} a_{\dot{C}} a_{\dot{D}}
\end{equation}
where the key function $W$ satisfies the expanding $\mathcal{HH}$ equation (\ref{expanding_HH_equation}) and $\Psi \ne 0$ is some function. It is obvious that the spinors $a_{A} = (0,1)$ and $a_{\dot{A}}$ are the 4-fold undotted and dotted Penrose spinors, respectively.

The expanding congruence of SD null strings consists of the integral manifolds of the Pfaff system 
\begin{equation}
\label{SD_Pfaff_system}
a_{A} g^{A \dot{A}} = 0 , \ \dot{A} = \dot{1}, \dot{2}
\end{equation}
Taking into account that $a_{A} = (0,1)$ and employing (\ref{PRF_tetrad}) one quickly infers that the SD null strings defined by (\ref{SD_Pfaff_system}) are 2-surfaces
\begin{equation}
\label{rownania_strun_SD_w_postaci_const}
q^{\dot{A}} = \textrm{const},  \ \dot{A} = \dot{1}, \dot{2}
\end{equation}
Consequently $(p^{\dot{1}}, p^{\dot{2}})$ are the coordinates on these null strings and any vector field tangent to such a null string has the form
\begin{equation}
\label{V+}
V^{(+)} = V^{(+) \dot{A}} \frac{\partial}{\partial p^{\dot{A}}}
\end{equation}
where $V^{(+) \dot{A}} = V^{(+) \dot{A}} (p^{\dot{B}})$.

Then the congruence of ASD null strings consists of the integral manifolds of the Pfaff system 
\begin{equation}
\label{ASD_Pfaff_system}
a_{\dot{A}} g^{A \dot{A}} = 0, \ A=1,2
\end{equation}
where $a_{\dot{A}}$ is the dotted 4-fold Penrose spinor given by (\ref{rozklad_ASD_krzywizny_na_spinory}). From (\ref{ASD_Pfaff_system}) with (\ref{PRF_tetrad}) we easily find that any vector field tangent to the ASD null string can be written as
\begin{equation}
\label{V-}
V^{(-)} = (na_{\dot{C}} Q^{\dot{A}\dot{C}}+ m a^{\dot{A}} ) \frac{\partial}{\partial p^{\dot{A}}} + n a^{\dot{A}} \frac{\partial}{\partial q^{\dot{A}}}
\end{equation}
where $n$ and $m$ are some functions on the ASD null string and $Q^{\dot{A}\dot{C}}$ is given by (\ref{postac_QAB}). Any SD null string and ASD null string intersect one another along a null geodesic line. The vector field $V$ tangent to this line must be of the form (\ref{V+}) and also of the form (\ref{V-}). So any vector field tangent to the null geodesic along the intersection of SD and ASD null string is of the form
\begin{equation}
\label{postac_wektora_stycznego_do_obu_strun}
V \sim a^{\dot{A}} \frac{\partial}{\partial p^{\dot{A}}}
\end{equation}
Intersections of the congruence of SD null strings with the congruence of ASD null strings constitute a congruence of null geodesics. The 1-form
\begin{equation}
\label{twist_three_form}
e := a_{\dot{A}} dq^{\dot{A}}
\end{equation}
annihilates all the vectors $V^{(+)}$, $V^{(-)}$ and $V$
\begin{equation}
e(V^{(+)}) = 0 , \ e(V^{(-)}) = 0 , \ e(V)=0
\end{equation}
Therefore, the Pfaff equation
\begin{equation}
\label{Pfaff_system}
e=0
\end{equation}
defines the complex 3-dimensional distribution $\mathcal{D}_{3}$ on $\mathcal{HH}$ space such that the SD distribution determined by the Pfaff system (\ref{SD_Pfaff_system}) and the ASD distribution determined by (\ref{ASD_Pfaff_system}) belong to $\mathcal{D}_{3}$. 

Define the \textsl{twist 3-form} $\mathcal{T}$ as
\begin{equation}
\label{twist_form}
\mathcal{T} = e \wedge de
\end{equation}
We say that the congruence of null geodesic is \textsl{notwisting} if the twist 3-form $\mathcal{T}$ vanishes, we say that it is \textsl{twisting} if $\mathcal{T} \ne 0$. From the Frobenius theorem it follows that the congruence of null geodesics is nontwisting iff the distribution $\mathcal{D}_{3}$ defined by (\ref{Pfaff_system}) is completely integrable \cite{56,57}.

Equivalently, one can define the scalar $2 \omega^{2} :=  \nabla_{[a} V_{b]} \, \nabla^{a} V^{b}$ with $V^{a}$ being the vector field tangent to the congruence of null geodesics. This scalar we call the \textsl{twist of congruence}. Analogous object plays an important role in geometrical optics of gravitational waves \cite{7,8,9,11,53,54,55}. Then we can easily show that the twist 3-form (\ref{twist_form}) vanishes at some point if and only if $\omega$ is equal zero at this point.

In this paper we consider the case of twisting congruence. Consequently, inserting (\ref{twist_three_form}) into (\ref{twist_form}), we assume that
\begin{equation}
\label{twist}
\mathcal{T} = \frac{1}{2} a^{\dot{B}} \frac{\partial a_{\dot{B}}}{\partial p^{\dot{C}}} dp^{\dot{C}} \wedge dq_{\dot{M}} \wedge dq^{\dot{M}} \ne 0 
\end{equation}
It can be easily demonstrated that if both SD and ASD congruences of null strings are nonexpanding then the respective congruence of null geodesic is necessary nontwisting. Hence, to get a twisting congruence of null geodesics one must assume that at least one of the congruences of null strings is expanding. Without any loss of generality one can assume that this is the congruence of SD null strings, and such a convention is accepted in the present paper.

We end the brief review of the $\mathcal{HH}$ formalism by introducing a useful coordinates called the \textsl{Plebański - Robinson - Finley (PRF) coordinates}. First we choose a spinor basis $(J_{\dot{A}}, K_{\dot{B}})$ where $J_{\dot{A}}$ is the constant spinor introduced before to define $\phi$, and $K_{\dot{A}}$ is another, independent of $J_{\dot{A}}$, constant spinor. Denote 
\begin{equation}
\label{definicja_tau}
\tau := K^{\dot{A}} J_{\dot{A}} \ne 0
\end{equation}
Then the PRF coordinates $(\eta, \phi,w,t)$ are defined by the formulae
\begin{equation}
\tau p^{\dot{A}} = \eta J^{\dot{A}} + \phi K^{\dot{A}} , \ \tau q^{\dot{A}} = t J^{\dot{A}} + w K^{\dot{A}}
\end{equation}
Using (\ref{definicja_tau}) we get
\begin{equation}
\eta = K^{\dot{A}} p_{\dot{A}} , \ \phi = J_{\dot{A}} p^{\dot{A}} , \ w = J_{\dot{A}} q^{\dot{A}} , \ t=  K^{\dot{A}} q_{\dot{A}}
\end{equation}
and, then
\begin{equation}
\label{wzory_na_pochodne_operatory}
\frac{\partial}{\partial p^{\dot{A}}} = -K_{\dot{A}} \frac{\partial}{\partial \eta} + J_{\dot{A}} \frac{\partial}{\partial \phi} , \ 
\frac{\partial}{\partial q^{\dot{A}}} = -K_{\dot{A}} \frac{\partial}{\partial t} + J_{\dot{A}} \frac{\partial}{\partial w} 
\end{equation}
The $\mathcal{HH}$ equation (\ref{expanding_HH_equation}) in terms of the PRF coordinates reads
\begin{equation}
\label{rownanie_HH}
   W_{\eta \eta}W_{\phi \phi} - W_{\eta \phi}^{2} + 2 \phi^{-1} W_{\eta} W_{\eta \phi} - 2 \phi^{-1} W_{\phi}W_{\eta\eta}  + (\tau \phi)^{-1} ( W_{w\eta}-W_{t\phi} )
 =  \frac{\varkappa}{2 \tau^{2}} \phi +  \frac{ \gamma}{\tau^{2}}
\end{equation}
and the condition (\ref{warunki_na_niezerowa_krzywizne}) takes the form
\begin{equation}
\label{warunki_na_niezerowa_krzywizne_2}
C_{2222} \ne 0 \ \Longleftrightarrow \ \gamma_{t} \ne 0
\end{equation}
Finally the metric (\ref{metryka_forma_1}) is
\begin{eqnarray}
\label{metryka_typu_Nxany}
ds^{2} &=& 2\phi^{-2} \big\{ \tau^{-1} (d \eta  d w - d \phi  dt) -    \phi \, W_{\eta\eta} \,   
dt^{2} \ \ \ \ \ \ 
\\ \nonumber
&& \ \ \ \ \ \ \ \ \ \ + ( 2W_{\eta} - 2\phi \, W_{\eta \phi}  ) \, dw dt
  +  ( 2  W_{\phi}  - \phi \, W_{\phi \phi}  ) \, dw^{2} \big\} \ \ \ \ \ \ 
\end{eqnarray}
Note that without any loss of generality one can put $\tau=1$.


\setcounter{equation}{0}
\section{New coordinates for the type $[\textrm{N}] \otimes [\textrm{N}]$, key function and $\mathcal{HH}$ equation}
\label{section_new_coordinates}

As was pointed out in previous section the expanding $\mathcal{HH}$ equation (\ref{rownanie_HH}) under the condition (\ref{warunki_na_niezerowa_krzywizne_2}) defines the key function $W$ for the type $[\textrm{N}] \otimes [\textrm{any}]$ with expanding congruence of SD null strings. Now we are going to specify the type of ASD part of the Weyl tensor to the type [N] and we will also demand that the congruence of null geodesics generated by the intersections of SD null strings with ASD null strings be twisting. From the previous section we conclude that to this end we must assume that (\ref{rozklad_ASD_krzywizny_na_spinory}) and (\ref{twist}) are fulfilled. In particular from (\ref{twist}) it follows that the 4-fold dotted Penrose spinor $a_{\dot{A}}$ cannot be proportional to any constant spinor. Consequently, one can always choose (locally) $a_{\dot{A}}$ as $a_{\dot{A}} = hJ_{\dot{A}} - K_{\dot{A}}$ or $a_{\dot{A}} = J_{\dot{A}} - h K_{\dot{A}}$, where $h=h(\eta, \phi,w,t)$ is some function such that $|h_{\eta}| + |h_{\phi}| \ne 0$. Further on we assume that
\begin{equation}
\label{definition_of_spinor_a}
a_{\dot{A}} = hJ_{\dot{A}} - K_{\dot{A}}
\end{equation}
Inserting (\ref{definition_of_spinor_a}) into (\ref{rozklad_ASD_krzywizny_na_spinory}) and employing (\ref{wzory_na_pochodne_operatory}) one arrives at the relations
\begin{equation}
\label{wzor_na_Psi}
\Psi = W_{\eta \eta \eta \eta} \ne 0
\end{equation}
and 
\begin{eqnarray}
\label{rownania_na_typ_N_ASD}
&& W_{\eta \eta \eta \phi} = h W_{\eta \eta \eta \eta}
\\ \nonumber
&& W_{\eta \eta \phi \phi} = h W_{\eta \eta \eta \phi}
\\ \nonumber
&& W_{\eta \phi \phi \phi} = h W_{\eta \eta \phi \phi}
\\ \nonumber
&& W_{\phi \phi \phi \phi} = h W_{\eta \phi \phi \phi}
\end{eqnarray}
Substituting (\ref{wzor_na_Psi}) into (\ref{rozklad_ASD_krzywizny_na_spinory}) we get from (\ref{ASD_conformal_curvature})
\begin{equation}
\label{ASD_curvature_written_byA}
C_{\dot{A}\dot{B}\dot{C}\dot{D}} = \phi^{3} W_{\eta \eta \eta \eta} a_{\dot{A}} a_{\dot{B}} a_{\dot{C}} a_{\dot{D}}
\end{equation}
Note that Eqs. (\ref{rownania_na_typ_N_ASD}) are equivalent to the following system of three Monge - Ampere equations
\begin{eqnarray}
\label{Monge_Ampere_system}
&& W_{\eta \eta \eta \eta} W_{\eta \eta \phi \phi} - W_{\eta \eta \eta \phi}^{2} = 0
\\ \nonumber
&& W_{\eta \eta \eta \phi} W_{\eta \phi \phi \phi} - W_{\eta \eta \phi \phi}^{2} = 0
\\ \nonumber
&& W_{\eta \eta \phi \phi} W_{\phi \phi \phi \phi} - W_{\eta \phi \phi \phi}^{2} = 0
\end{eqnarray}
Differentiating the first of Eqs. (\ref{rownania_na_typ_N_ASD}) with respect to $\phi$ and the second of Eqs. (\ref{rownania_na_typ_N_ASD}) with respect to $\eta$ and substracting the results one obtains the equation
\begin{equation}
\label{rownanie_na_h}
h_{\phi} - h h_{\eta} = 0
\end{equation}
Proceeding analogously with other equations of the system (\ref{rownania_na_typ_N_ASD}) we arrive at the same equation (\ref{rownanie_na_h}) which therefore appears to be the integrability condition of the system (\ref{rownania_na_typ_N_ASD}). We look for the solutions of Eq. (\ref{rownanie_na_h}) such that $h_{\eta} \ne 0$. The general solution of (\ref{rownanie_na_h}) is given by
\begin{equation}
\label{rozwiazanie_na_h}
\eta + \phi h = f(h,w,t)
\end{equation}
where $f=f(h,w,t)$ is an arbitrary function of its arguments. From (\ref{rozwiazanie_na_h}) one can extract the solution $h=h(\eta, \phi,w,t)$. The next step is to solve the system (\ref{rownania_na_typ_N_ASD}) or, equivalently, (\ref{Monge_Ampere_system}). Employing the general solution of the Monge - Ampere equation in the form given by D.B. Fairlie and A.N. Leznov in \cite{58} we can easily find $W_{\eta \eta \eta}$ and $W_{\eta \eta \phi}$ from the first of Eqs. (\ref{Monge_Ampere_system}). Then using the relations (\ref{rownania_na_typ_N_ASD}) one gets the remaining $W_{\eta \phi \phi}$ and $W_{\phi \phi \phi}$. Gathering, we have the following result
\begin{eqnarray}
\label{trzecie_pochodne}
&& W_{\eta \eta \eta} = \dddot{F}
\\ \nonumber
&& W_{\eta \eta \phi} = h \dddot{F} - \ddot{F}
\\ \nonumber
&& W_{\eta \phi \phi} = h^{2} \dddot{F} - 2h \ddot{F} + 2 \dot{F}
\\ \nonumber
&& W_{\phi \phi \phi} = h^{3} \dddot{F} - 3h^{2} \ddot{F} + 6h \dot{F} - 6F
\end{eqnarray}
where $F=F(h,w,t)$ is an arbitrary function of its arguments and the overdot denotes differentiation with respect to $h$ i.e., $\dot{F} := \dfrac{\partial F}{\partial h}$, $\ddot{F} := \dfrac{\partial^{2} F}{\partial h^{2}}$, ... etc. From (\ref{trzecie_pochodne}) we have 
\begin{eqnarray}
&& W_{\eta \eta \eta \eta} = h_{\eta} \ddddot{F} \ , \ \ \ W_{\eta \eta \eta \phi} = hh_{\eta} \ddddot{F} \ , \ \ \ 
W_{\eta \eta \phi \phi} = h^{2}h_{\eta} \ddddot{F}
\\ \nonumber
&& W_{\eta \phi \phi \phi} = h^{3}h_{\eta} \ddddot{F} \ , \ \ \ W_{\phi \phi \phi \phi} = h^{4}h_{\eta} \ddddot{F}
\end{eqnarray}
So (\ref{ASD_curvature_written_byA}) can be rewritten as
\begin{equation}
\label{ASD_krzywizna_z_F}
C_{\dot{A}\dot{B}\dot{C}\dot{D}} = \phi^{3} h_{\eta} \ddddot{F} a_{\dot{A}} a_{\dot{B}} a_{\dot{C}} a_{\dot{D}}
\end{equation}
and we infer that necessary
\begin{equation}
\label{warunek_na_niezerowosc_ddddotF}
\ddddot{F} \ne 0
\end{equation}
Equations (\ref{trzecie_pochodne}) can be integrated and this gives us the key function $W$ in the form
\begin{equation}
\label{funkcja_kluczowa}
W = -F \phi^{3} + \frac{1}{2} (R - 2hS + h^{2} \Omega) \phi^{2} + (B-Ah) \phi +C
\end{equation}
where $F,R,S, \Omega, B,A$ and $C$ are functions depending on $(h,w,t)$ satisfying the following relations
\begin{eqnarray}
\label{zaleznosci_miedzy_funkcjami}
&& \dot{A} = \Omega \dot{f} \ , \ \ \ \dot{B} = S \dot{f} \ , \ \ \ \dot{C} = A \dot{f} \ , \ \ \ \dot{\Omega} = \dddot{F} \dot{f}
\\ \nonumber
&& \dot{S} = (h \dddot{F} - \ddot{F}) \dot{f} \ , \ \ \ \dot{R} = (h^{2} \dddot{F} - 2h \ddot{F} + 2\dot{F}) \dot{f}
\end{eqnarray}

[\textbf{Remark}: To perform the integration of Eqs. (\ref{trzecie_pochodne}) we intensively use the formulae which follow from Eq. (\ref{rozwiazanie_na_h}) by differentiating both sides of that equation with respect to $\eta$ or $\phi$. Thus we get
\begin{eqnarray}
\label{wniosek_z_postaci_h}
&& 1+\phi h_{\eta} = \dot{f} h_{\eta}  \ \Longrightarrow \ h_{\eta} = \frac{1}{\dot{f}-\phi} 
\\ \nonumber
&& h+\phi h_{\phi} = \dot{f} h_{\phi}  \ \Longrightarrow \ h_{\phi} = \frac{h}{\dot{f}-\phi}
\end{eqnarray}
Analogously one quickly finds the relations 
\begin{eqnarray}
\label{drugi_wniosek_z_postaci_h}
&& \phi h_{t} = \dot{f} h_{t}  + f_{t} \ \Longrightarrow \ h_{t} = \frac{f_{t}}{\phi-\dot{f}}
\\ \nonumber
&& \phi h_{w} = \dot{f} h_{w}  + f_{w} \ \Longrightarrow \ h_{w} = \frac{f_{w}}{\phi-\dot{f}}
\end{eqnarray}
where $f_{t} := \dfrac{\partial f(h,w,t)}{\partial t}$ and $f_{w} := \dfrac{\partial f(h,w,t)}{\partial w}$. Eqs. (\ref{wniosek_z_postaci_h}) and (\ref{drugi_wniosek_z_postaci_h}) are also useful when the $\mathcal{HH}$ equation in new coordinates $(h,\phi,w,t)$ is considered and when the master equation for a homothetic symmetry is investigated (see Sec. \ref{section_Homothetic_symmetries})].

Exterior differential of (\ref{rozwiazanie_na_h}) multiplied by $d\phi \wedge dw \wedge dt$ leads to the relation
\begin{equation}
d\eta \wedge d\phi \wedge dw \wedge dt = (\dot{f}- \phi) dh \wedge d\phi \wedge dw \wedge dt
\end{equation}
Since $d\eta \wedge d\phi \wedge dw \wedge dt \ne 0$ one infers that $\dot{f} - \phi \ne 0$ (see (\ref{wniosek_z_postaci_h}) and (\ref{drugi_wniosek_z_postaci_h})) and
\begin{equation}
dh \wedge d\phi \wedge dw \wedge dt \ne 0
\end{equation}
Therefore, we can introduce new coordinates $(h, \phi,w,t)$. The obvious advantage of such a choice is the fact that according to (\ref{funkcja_kluczowa}) the key function $W$ expressed as a function of $(h, \phi,w,t)$ appears to be the third-order polynomial in $\phi$ with coefficients dependent on $(h,w,t)$. 

At this point a comment is needed. J.D. Finley and J.F. Plebański motivated by the results of \cite{48} also arrived at the conclusion that the key function is a third-order polynomial in $\phi$ if some new coordinate $h$ is used \cite{20,24}. The coordinate $h$ employed in \cite{20,24} is defined in other way than "our" $h$. So to avoid confusion we denote the $h$ introduced by Finley and Plebański by $\tilde{h}$. Then one can quickly show that the relation between $\tilde{h}$ and the $h$ used in the present paper reads
\begin{equation}
\label{zwiazek_miedzy_h_i_tildeh}
h=\frac{\tilde{h}_{\phi}}{\tilde{h}_{\eta}} , \  \tilde{h} = \tilde{h} (\eta , \phi, w,t)
\end{equation}
Employing (\ref{rownanie_na_h}) we show that the solution of (\ref{zwiazek_miedzy_h_i_tildeh}) is any function $\tilde{h}$ of the form
\begin{equation}
\tilde{h} = \tilde{h} (h,w,t) , \ \tilde{h}_{h} \ne 0
\end{equation}
where, to remind of this, $h$ is defined by (\ref{definition_of_spinor_a}) or, equivalently, by (\ref{rownania_na_typ_N_ASD}). 

From (\ref{rownania_strun_SD_w_postaci_const}), (\ref{postac_wektora_stycznego_do_obu_strun}), (\ref{definition_of_spinor_a}) and (\ref{rownanie_na_h}) one infers that the congruence of null geodesics consisting of the intersections of SD null strings with ASD null strings is defined by the equations 
\begin{equation}
w = \textrm{const} , \ t=\textrm{const} , \ h=\textrm{const}
\end{equation}
Returning now to the form of key function $W$ (\ref{funkcja_kluczowa}) in variables $(h, \phi,w,t)$ we insert this function into $\mathcal{HH}$ equation (\ref{rownanie_HH}). Then employing (\ref{wniosek_z_postaci_h}) and (\ref{drugi_wniosek_z_postaci_h}), and comparing the terms standing at the same powers of $\phi$ one arrives at the \textsl{$\mathcal{HH}$ system of equations}
\begin{subequations}
\label{uklad_rownan_na_typN_bez_symetrii}
\begin{eqnarray}
\label{uklad_rownan_na_typN_bez_symetrii_1}
&& (R+ h^{2} \Omega -2hS) \ddot{F}  +2(S-h\Omega) \dot{F} + \frac{1}{\tau} ( \dot{F}_{w}- h \dot{F}_{t} +3F_{t} ) = \frac{\varkappa}{2 \tau^{2}}
\\ 
\label{uklad_rownan_na_typN_bez_symetrii_2}
&& S^{2}-\Omega R   + 4A \dot{F} + 2(B- hA  ) \ddot{F} + \frac{1}{\tau} \left( hS_{t}- R_{t}   - f_{t} (h \ddot{F}-2 \dot{F})  \right. 
\\ \nonumber
&& \ \ \ \ \ \ \ \ \ \ \left. + S_{w} - h \Omega_{w} + f_{w} \ddot{F}  \right) = \frac{\gamma}{\tau^{2}} 
\\ 
\label{uklad_rownan_na_typN_bez_symetrii_3}
&& 2SA-2\Omega B + \frac{1}{\tau} \left( A_{w} - f_{w} \Omega - B_{t} + f_{t}S   \right) = 0
\end{eqnarray}
\end{subequations}
where the functions $F,R,S, \Omega, B,A$ and $f$ are considered as dependent on $(h,w,t)$, so $A_{t} \equiv \dfrac{\partial A(h,w,t)}{\partial t}$, $A_{w} \equiv \dfrac{\partial A(h,w,t)}{\partial w}$,..., etc. Since by (\ref{zaleznosci_miedzy_funkcjami}) the functions $F$ and $f$ can be handled as independent functions and all remaining unknown functions are expressed by the integrals over $h$ with integrands determined by $F$ and $f$. We conclude that Eqs. (\ref{uklad_rownan_na_typN_bez_symetrii}) constitute the system of three equations for two functions $F$ and $f$ and, consequently, in general this system is overdetermined. In the next section we assume that the $\mathcal{HH}$ space admits two homothetic vectors and we will see later that presence of such two symmetries enables us to reduce the system (\ref{uklad_rownan_na_typN_bez_symetrii}) to one ODE for a single function.


\setcounter{equation}{0}
\section{Homothetic symmetries}
\label{section_Homothetic_symmetries}

A projective vector field $X$ is called the \textsl{homothetic vector field} (or simply the \textsl{homothetic vector}) if the Lie derivative of the metric tensor $g_{ab}$ along $X$ satisfies the condition
\begin{equation}
\pounds_{X} g_{ab} = \chi g_{ab}
\end{equation}
where $\chi = \textrm{const}$. If $\chi = \textrm{const} \ne 0$ then $X$ is called the \textsl{proper homothetic vector field} (the \textsl{proper homothetic vector}); if $\chi=0$ then $X$ is called the \textsl{Killing vector field} (the \textsl{Killing vector}).

Assume that $X$ and $Y$ are two independent proper homothetic vectors
\begin{equation}
\label{dwa_homotetyczne_Killingi}
\pounds_{X} g_{ab} = \chi_{1} g_{ab} , \ \pounds_{Y} g_{ab} = \chi_{2} g_{ab}, \ \chi_{1}, \chi_{2} = \textrm{constants} \ne 0
\end{equation}
Define two independent vectors
\begin{equation}
\tilde{X} := \frac{\chi_{2}}{\chi_{1}} X - Y , \ \tilde{Y} := Y
\end{equation}
By (\ref{dwa_homotetyczne_Killingi}) we have
\begin{equation}
\pounds_{\tilde{X}} g_{ab} = 0 , \ \pounds_{\tilde{Y}} g_{ab} = \chi_{2} g_{ab}
\end{equation}
So $\tilde{X}$ is the Killing vector and $\tilde{Y}$ is the proper homothetic vector. Therefore one concludes that if the space admits two linearly independent homothetic vectors then without any loss of generality we can assume that, at least, one of them is a Killing vector.

Return now to our $\mathcal{HH}$ space of the type $[\textrm{N}] \otimes [\textrm{N}]$ with expanding congruence of SD null strings and with twisting congruence of null geodesics generated by intersections of SD and ASD null strings. Moreover, we assume that this $\mathcal{HH}$ space admits two linearly independent homothetic vector fields. As has just been pointed out one of them, say $K_{1}$, is the Killing vector and the second, denoted by $K_{2}$ is the homothetic vector, i.e.
\begin{equation}
\label{rownania_Killinga}
\pounds_{K_{1}} g_{ab} = 0 , \ \pounds_{K_{2}} g_{ab} = \chi_{0} g_{ab} , \ \chi_{0} = \textrm{const}
\end{equation}
Careful analysis of the results found by one of us (A. Ch.) in \cite{59} (see also the work by A. Sonnleitner and J.D. Finley \cite{60} and by J.D. Finley \cite{38}) leads to the conclusion, that there exist the PRF coordinates $(\eta, \phi,w,t)$ such that the Killing vector takes the form
\begin{equation}
\label{Killing_K1}
K_{1} = \frac{\partial}{\partial w}
\end{equation}
the key function $W$ is independent of $w$
\begin{equation}
\label{funkcja_kluczowa_z_jednym_Killingiem}
W=W(\eta,\phi,t)
\end{equation}
and $\varkappa = \varkappa (t)$, $\gamma = \gamma (t)$, $\gamma_{t} \ne 0$.

Then, still employing the results of \cite{59} we infer that the homothetic vector $K_{2}$ must be of the form
\begin{equation}
\label{Killing_ogolnie}
K_{2} = a \frac{\partial}{\partial w} + b \frac{\partial}{\partial t} + (b_{t}-2\chi_{0}) \phi \frac{\partial}{\partial \phi} +[(2b_{t}-a_{w}-2\chi_{0}) \eta + b_{w} \phi - \tau \epsilon ] \frac{\partial}{\partial \eta}
\end{equation}
Moreover, ten equations for $K_{2}$ (\ref{rownania_Killinga}) and their integrability conditions can be brought to the \textsl{master equation}
\begin{eqnarray}
\label{master_equation}
\pounds_{K_{2}} W &=& - (4 \chi_{0} + 2 a_{w} - 3 b_{t}) W + \alpha \phi^{3}  + \frac{1}{2} (\epsilon_{w} \phi + \epsilon_{t} \eta) + \beta
\\ \nonumber
&& + \frac{1}{2 \tau} [-b_{ww} \phi^{2} - b_{tt} \eta^{2} + (a_{ww} - 2b_{tw}) \eta \phi ]
\end{eqnarray}
and to the constraint equation
\begin{equation}
\label{integrability_condition_of_master_equation}
b = \frac{a_{www} - 4a_{w} \gamma}{2 \gamma_{t}}
\end{equation}
where $a=a(w)$, $b=b(w,t)$, $\epsilon = \epsilon (w,t)$, $\alpha = \alpha (w,t)$ and $\beta = \beta (w,t)$ are some functions of their arguments.

A crucial point is to note also that we have some coordinate gauge freedom and a gauge freedom for the key function $W$ to our disposal. This gauge freedom and the corresponding transformations of the objects $a,b,\epsilon, \varkappa, \gamma, \alpha$ and $\beta$ read (see \cite{59} for the details)
\begin{eqnarray}
\label{gauge_freedom}
&& w'=w+w_{0}, \ w_{0} = \textrm{const}, \ t'=t'(t) , \ t'_{t} =: \lambda^{-\frac{1}{2}}
\\ \nonumber
&& \phi'=\lambda^{-\frac{1}{2}} \phi , \ \eta' = \lambda^{-1} \eta + \tau \sigma 
\\ \nonumber
&& W' = \lambda^{-\frac{3}{2}} W - \frac{1}{2 \tau} \lambda^{-1} (\lambda^{-\frac{1}{2}})_{t} \eta^{2} - \frac{1}{2} \lambda^{-\frac{1}{2}} \sigma_{t} \eta - \frac{1}{3} \lambda^{-\frac{3}{2}} L \phi^{3} - \lambda^{-\frac{3}{2}} M 
\\ \nonumber
&& a'=a , \ b' = \lambda^{-\frac{1}{2}} b , \ \epsilon' = \lambda^{-1} \epsilon - \sigma (2\chi_{0} + a_{w} - 2b_{t} + b (\ln \sigma \lambda)_{t})
\\ \nonumber
&& \varkappa' = \lambda^{\frac{1}{2}} \varkappa + 2 \tau \lambda^{\frac{1}{2}} L_{t} , \ \gamma'=\gamma , \ \gamma'_{t'} = \lambda^{\frac{1}{2}} \gamma_{t}
\\ \nonumber
&& \alpha' = \alpha - \frac{1}{3} b L_{t} + \frac{2}{3} (\chi_{0} - a_{w}) L
\\ \nonumber
&& \beta' = \lambda^{-\frac{3}{2}} \beta  - \frac{1}{2} \tau \sigma \lambda^{\frac{1}{2}} \epsilon'_{t} + \frac{1}{2} \tau \lambda^{\frac{1}{2}} \sigma^{2} [\lambda^{\frac{1}{2}} (b \lambda^{-\frac{1}{2}})_{t}]_{t} + \frac{1}{2} \tau \lambda^{-\frac{1}{2}} \epsilon \sigma_{t} 
\\ \nonumber
&& \ \ \ \ \ \  - \lambda^{-\frac{3}{2}} (4 \chi_{0} - 3 b_{t} + 2 a_{w}) M - \lambda^{-\frac{3}{2}} b M_{t}
\end{eqnarray}
where 
$t'(t)$, $\lambda = \lambda(t)$, $\sigma = \sigma (t)$, $L=L(t)$ and $M=M(t)$ are arbitrary gauge functions. 

Another important fact which should be taken into account when we are looking for the homothetic vector $K_{2}$ is that by the Yano theorem \cite{38,61,62}: \textsl{the commutator of any two homothetic vectors, if it does not vanish, is a Killing vector}. 

Employing this theorem and the gauge freedom (\ref{gauge_freedom}), and performing a bit laborious analysis of the master equation (\ref{master_equation}) and the constraint equation (\ref{integrability_condition_of_master_equation}) one arrives at two important conclusions:

First, the homothetic vector $K_{2}$ can be brought to the form
\begin{equation}
\label{Killing_K2}
K_{2} = w \frac{\partial}{\partial w} + t \frac{\partial}{\partial t} + (1-2\chi_{0}) \left( \phi \frac{\partial}{\partial \phi} + \eta \frac{\partial}{\partial \eta} \right)
\end{equation}
(with the Killing vector $K_{1}$ given by (\ref{Killing_K1})).

The commutator of $K_{1}$ and $K_{2}$ is
\begin{equation}
[K_{1}, K_{2}]=K_{1}
\end{equation}
Second, \textsl{a $\mathcal{HH}$ space of the type $[\textrm{N}] \otimes [\textrm{N}]$ with expanding congruence of SD null strings and twisting congruence of null geodesics generated by the intersections of SD and ASD null strings does not admit a group of homothetic motions of order $>2$}. (This is a "complex" version of the results obtained in \cite{16,17}).

From (\ref{integrability_condition_of_master_equation}) and (\ref{Killing_K2}) one quickly gets
\begin{equation}
\label{rozwiazanie_na_gamma}
\gamma (t) = \gamma_{0} t^{-2}, \ \gamma_{0} = \textrm{const} \ne 0
\end{equation}
Inserting (\ref{funkcja_kluczowa_z_jednym_Killingiem}) into (\ref{rownania_na_typ_N_ASD}) we conclude that the function $h$ is independent of $w$
\begin{equation}
h=h(\eta, \phi,t)
\end{equation}
Consequently, the function $f$ introduced in Eq. (\ref{rozwiazanie_na_h}) is also independent of $w$
\begin{equation}
f=f(h,\phi,t)
\end{equation}
Assume first that
\begin{equation}
\label{warunek_niezerowosci_dotf}
\frac{\partial f}{\partial h} \equiv \dot{f} \ne 0
\end{equation}
Considering the key function $W$ as dependent on $(h,\phi,t)$ according to (\ref{funkcja_kluczowa}), then inserting it into the master equation (\ref{master_equation}) with $K_{2}$ given by (\ref{Killing_K2}) and comparing the terms standing at the same powers of $\phi$ we get the system of differential equations which can be easily solved under the conditions (\ref{zaleznosci_miedzy_funkcjami}) and (\ref{warunek_niezerowosci_dotf}). The solution can be presented in the form 
\begin{eqnarray}
\label{rozwiazania_po_uwaglednieniu_K2}
&& \alpha=0, \ \beta = 0, \ F(h,t) = t^{2(\chi_{0}-1)} u(h) , \ f(h,t) = t^{1-2\chi_{0}} v(h)
\\ \nonumber
&& C(h,t) = t^{1-4\chi_{0}} N(v), \ A(h,t) = t^{-2\chi_{0}} \frac{dN}{dv} , \ B(h,t) = t^{-2\chi_{0}} Z(v)
\\ \nonumber
&& \Omega (h,t) = t^{-1} \frac{d^{2} N}{dv^{2}} , \ R(h,t) = t^{-1} P(v) , \ S(h,t) = t^{-1} \frac{dZ}{dv}
\end{eqnarray}
where $u=u(h)$, $v=v(h)$, $N=N(v) = N(v(h))$, $Z = Z(v) = Z(v(h))$, $P=P(v) = P(v(h))$ are functions of their arguments. The functions $N$, $P$ and $Z$ are related to the function $u$ as follows
\begin{eqnarray}
\label{zwiazki_U_z_NZP}
&& \frac{d^{3} N}{d v^{3}} = \dddot{u}  
\\ \nonumber
&& \frac{d^{2} Z}{d v^{2}} = h \dddot{u}  - \ddot{u} 
\  \ \ \ \ \ \ \ \ \ \ \ \ \ \ \ \  \Longrightarrow \ \ \ \ \ \ddot{u} = h \frac{d^{3} N}{d v^{3}} - \frac{d^{2} Z}{d v^{2}}
\\ \nonumber
&& \frac{d P}{d v} = h^{2} \dddot{u}  - 2h \ddot{u} + 2 \dot{u}
\  \ \ \ \ \  \Longrightarrow \ \ \ \ \
2 \dot{u} = \frac{d P}{d v} - 2h \frac{d^{2} Z}{d v^{2}} + h^{2} \frac{d^{3} N}{d v^{3}}
\end{eqnarray}
Before we substitute (\ref{rozwiazania_po_uwaglednieniu_K2}) and (\ref{zwiazki_U_z_NZP}) into the system of equations (\ref{uklad_rownan_na_typN_bez_symetrii}) we should study the master equation (\ref{master_equation}) under (\ref{Killing_K2}) in the opposite case to that defined by (\ref{warunek_niezerowosci_dotf}). So now we assume that
\begin{equation}
\label{warunek_zerowosci_dotf}
\dot{f} =0 \ \Longrightarrow f=f(t)
\end{equation}
First, from (\ref{rownania_na_typ_N_ASD}) and (\ref{rozwiazanie_na_h}) one quickly finds that the gauge transformation (\ref{gauge_freedom}) induces the following transformations of $h$ and $f$
\begin{eqnarray}
&& h' = \frac{W'_{\eta' \eta' \eta' \phi'}}{W'_{\eta' \eta' \eta' \eta'}} = \frac{W_{\eta \eta \eta \phi}}{W_{\eta \eta \eta \eta}} \frac{\eta'_{\eta}}{\phi'_{\phi}} = \lambda^{-\frac{1}{2}} h
\\ \nonumber
&& f' = \eta' + \phi' h' = \lambda^{-1} f + \tau \sigma
\end{eqnarray}
Therefore it is evident that one can always choose $\sigma(t)$ such that $f'=0$. Consequently, if (\ref{warunek_zerowosci_dotf}) holds true then without any loss of generality one can put
\begin{equation}
\label{warunek_zerowosci_f}
f=0
\end{equation} 
and this inserted into (\ref{rozwiazanie_na_h}) gives a simple formula for $h$
\begin{equation}
h = - \frac{\eta}{\phi}
\end{equation}
Now the crucial question arises whether the $\sigma$-gauge which brings $f$ to zero does not change the form of the homothetic vector (\ref{Killing_K2}) by making $\epsilon \ne 0$ (see (\ref{Killing_ogolnie}) and the transformation of $\epsilon$ in (\ref{gauge_freedom})). To answer this question observe that substituting (\ref{warunek_zerowosci_f}) into (\ref{zaleznosci_miedzy_funkcjami}) we find that the functions $A,B,C,\Omega,S$ and $R$ are independent of $h$. Then taking $W$ as the function dependent on $(h,\phi,t)$ according to (\ref{funkcja_kluczowa}), and $a=w$, $b=t$ according to (\ref{Killing_K2}) one can easily solve the master equation (\ref{master_equation}) under the assumption (\ref{warunek_zerowosci_f}). The solution reads

\begin{eqnarray}
\label{rozwiazania_po_uwaglednieniu_K2_dla_f0}
&& \alpha=0, \ \beta = 0, \ \epsilon=0,\ F(h,t) = t^{2(\chi_{0}-1)} u(h) 
\\ \nonumber
&& C(t) = C_{0} t^{1-4\chi_{0}}, \ A(t) =A_{0} t^{-2\chi_{0}}  , \ B(t) = B_{0} t^{-2\chi_{0}} 
\\ \nonumber
&& \Omega (t) = \Omega_{0} t^{-1}  , \ R(t) = R_{0} t^{-1} , \ S(t) = S_{0} t^{-1} 
\\ \nonumber
&& A_{0}, B_{0}, C_{0}, \Omega_{0}, S_{0}, R_{0} = \textrm{const}
\end{eqnarray}
As can be seen $\epsilon=0$ and this means that the $\sigma$-gauge making $f=0$ maintains the function $\epsilon$ vanishing.

In the next two sections we investigate the \textsl{$\mathcal{HH}$ system of equations (\ref{uklad_rownan_na_typN_bez_symetrii})} when $\dot{f}=0$ and then when $\dot{f} \ne 0$.


\setcounter{equation}{0}
\section{The key function and the metric of twisting type $[\textrm{N}] \otimes [\textrm{N}]$ $\mathcal{HH}$ space with $\dot{f}=0$} 
\label{section_f_zero}

As it has been shown in the former section one can put here $f=0$. So our task is to investigate the $\mathcal{HH}$ system of equations (\ref{uklad_rownan_na_typN_bez_symetrii}) under the conditions (\ref{rozwiazania_po_uwaglednieniu_K2_dla_f0}), (\ref{warunek_zerowosci_f}) and (\ref{rozwiazanie_na_gamma}). Differentiating (\ref{uklad_rownan_na_typN_bez_symetrii_2}) twice with respect to $h$ one gets
\begin{equation}
\label{f0_wniosek_1}
(B(t) - hA(t)) \ddddot{F} = 0 \stackrel{(\ref{warunek_na_niezerowosc_ddddotF})}{\Longrightarrow}  A(t)=B(t)=0
\end{equation}
Then Eq. (\ref{uklad_rownan_na_typN_bez_symetrii_2}) reads now 
\begin{equation}
S_{0}^{2} - \Omega_{0} R_{0} + \frac{1}{\tau} (R_{0} - hS_{0}) = \frac{\gamma_{0}}{\tau^{2}}
\end{equation}
Hence
\begin{eqnarray}
\label{f0_wniosek_3}
&& S_{0} = 0 \Longrightarrow S(t)=0
\\ \nonumber
&& R_{0} (1-\tau \Omega_{0}) = \frac{\gamma_{0}}{\tau} \ne 0 \Longrightarrow \Omega_{0} \ne \tau^{-1} \ \textrm{and} \ R_{0} = \frac{\gamma_{0}}{\tau (1-\tau \Omega_{0})} \ne 0
\\ \nonumber
&& R(t) = \frac{\gamma_{0}}{\tau (1-\tau \Omega_{0})} \, t^{-1}
\end{eqnarray}
Obviously Eq. (\ref{uklad_rownan_na_typN_bez_symetrii_3}) is trivially satisfied and we are left with Eq. (\ref{uklad_rownan_na_typN_bez_symetrii_1}). Inserting into this equation the function $F(h,t)$ given in (\ref{rozwiazania_po_uwaglednieniu_K2_dla_f0}), and employing also (\ref{f0_wniosek_1}) and (\ref{f0_wniosek_3}) one gets
\begin{equation}
\label{rownanie_przy_f0}
(\Omega_{0} h^{2} + R_{0}) \ddot{u} - 2h  \left( \Omega_{0} + \frac{\chi_{0}-1}{\tau} \right) \dot{u} + \frac{6(\chi_{0}-1)}{\tau} u = \frac{\varkappa (t)}{2 \tau^{2}} t^{3-2\chi_{0}}
\end{equation}
The left hand side of (\ref{rownanie_przy_f0}) depends on $h$ only and the right hand side on $t$. Consequently, we must have
\begin{equation}
\label{rozwiazanie_na_varkappa}
\frac{\varkappa (t)}{2 \tau^{2}} t^{3-2\chi_{0}} =: F_{0} = \textrm{const}
\end{equation}
Finally, the $\mathcal{HH}$ system of equations (\ref{uklad_rownan_na_typN_bez_symetrii}) in the present case reduces to one equation 
\begin{equation}
\label{rownanie_przy_f0_2}
(\Omega_{0} h^{2} + R_{0}) \ddot{u} - 2 h \left( \Omega_{0} + \frac{\chi_{0}-1}{\tau} \right) \dot{u} + \frac{6(\chi_{0}-1)}{\tau} u -F_{0}=0
\end{equation}
with $R_{0} \ne 0$ and $\Omega_{0} \ne \tau^{-1}$. Then, the key function $W$ has now the form 
\begin{equation}
\label{funkcja_kluczowa_dla_przypadku_f0}
W = - u(h) \phi^{3} t^{2(\chi_{0}-1)} + \frac{1}{2} (R_{0} + h^{2} \Omega_{0}) \phi^{2} t^{-1}+C_{0}t^{1-4\chi_{0}}
, \ h=-\frac{\eta}{\phi} 
\end{equation}
with $u=u(h)$ satisfying the second order linear ODE (\ref{rownanie_przy_f0_2}).

Since $R_{0} \ne 0$ we can look for the solution of Eq. (\ref{rownanie_przy_f0_2}) in the form of power series
\begin{equation}
\label{propozycja_rozwiazania}
u= \sum_{n=0}^{\infty} c_{n} h^{n}
\end{equation}
Substituting (\ref{propozycja_rozwiazania}) into (\ref{rownanie_przy_f0_2}) and comparing the terms at the same power of $h$ one gets the following recurrent equation
\begin{equation}
\label{rekurencyjny_wzor}
c_{n+2} = \frac{-\left( \Omega_{0} n - \dfrac{2(\chi_{0}-1)}{\tau} \right)  (n-3) c_{n} +F_{0} \delta_{n0}}{R_{0}(n+2)(n+1)} , \ n=0,1,...
\end{equation}
From (\ref{rekurencyjny_wzor}) we quickly infer the formulae for the coefficients $c_{n}$
\begin{eqnarray}
&& c_{3} = \frac{\Omega_{0} - \dfrac{2(\chi_{0}-1)}{\tau}}{3 R_{0}} \, c_{1} , \ c_{2n+1} = 0 \ \textrm{for} \ n \geq 2
, \  c_{2} = \frac{F_{0} - \dfrac{6(\chi_{0}-1)}{\tau} \, c_{0}}{2R_{0}}
\\ \nonumber
&& c_{2(n+1)} = (-1)^{n+1} \frac{ \left( \Omega_{0} - \dfrac{\chi_{0}-1}{\tau} \right) \left( 2\Omega_{0} - \dfrac{\chi_{0}-1}{\tau} \right) ... \left(n \Omega_{0} - \dfrac{\chi_{0}-1}{\tau} \right) }{R_{0}^{n} (n+1)! (4n^{2}-1)} \,  c_{2}
\ \textrm{for} \ n \geq 1
\end{eqnarray}
Therefore, the series (\ref{propozycja_rozwiazania}) reads
\begin{equation}
\label{rozwiazanie_w_postaci_szeregu}
u(h) = c_{0} + c_{1} \left( h+ \frac{\Omega_{0} - \dfrac{2(\chi_{0}-1)}{\tau}}{3R_{0}} \, h^{3}  \right) + 
 c_{2} \left( h^{2} + \sum_{n=1}^{\infty} (-1)^{n+1}  a_{n} h^{2(n+1)}  \right)
\end{equation}
where
\begin{equation}
a_{n} := \dfrac{\left( \Omega_{0} - \dfrac{\chi_{0}-1}{\tau} \right) \left( 2\Omega_{0} - \dfrac{\chi_{0}-1}{\tau} \right) ... \left(n \Omega_{0} - \dfrac{\chi_{0}-1}{\tau} \right)}{R_{0}^{n} (n+1)! (4n^{2}-1)}
\end{equation}
This series reduces to a polynomial if there exists $n \in \{ 1,2,... \}$ such that
\begin{equation}
\label{warunek_na_wielomian}
n\Omega_{0} = \frac{\chi_{0}-1}{\tau}
\end{equation}
In the opposite case $u=u(h)$ is the infinite series of the radius of convergence $r_{c}$
\begin{equation}
r_{c} = \left| \frac{R_{0}}{\Omega_{0}} \right|
\end{equation}
and so, for $|h| < \left| \dfrac{R_{0}}{\Omega_{0}} \right|$ this series with $c_{0}$ and $c_{1}$ being arbitrary constants gives the general analytic solution of Eq. (\ref{rownanie_przy_f0_2}).

However, in order to ensure that our $\mathcal{HH}$ space is of the type $[\textrm{N}] \otimes [\textrm{N}]$ the coordinate $h$ should be also restricted by the condition
\begin{equation}
\label{conditions_na_niezerowy_twist}
 \ddddot{u} \ne 0
\end{equation}
which follows from (\ref{warunek_na_niezerowosc_ddddotF}) with (\ref{rozwiazania_po_uwaglednieniu_K2_dla_f0}). From (\ref{conditions_na_niezerowy_twist}) one easily infers that in the case when the series (\ref{rozwiazanie_w_postaci_szeregu}) reduces to the polynomial we should assume that (\ref{warunek_na_wielomian}) holds true with $n \geq 2$. Then the respective polynomial is of the order $2n$. Consequently, the minimal order of the polynomial $u(h)$ defined by (\ref{rozwiazanie_w_postaci_szeregu}) leading to the $[\textrm{N}] \otimes [\textrm{N}]$ twisting $\mathcal{HH}$ space admitting two homothetic vectors is four. It is an easy matter to note that such a class of  $\mathcal{HH}$ spaces belongs to the class investigated in our previous work \cite{42}.

A quick glance at the form of metric (\ref{metryka_typu_Nxany}) shows that this metric depends on the first and second derivatives of $u(h)$ and not on the function $u(h)$ itself. Define
\begin{eqnarray}
\label{wzor_na_pochodna_u}
&& U(h) := \frac{du}{dh}= c_{1} \left( 1+ \frac{\Omega_{0} - \dfrac{2(\chi_{0}-1)}{\tau}}{R_{0}} \, h^{2} \right)
\\ \nonumber
&& +2c_{2} \left( h + \sum_{n=1}^{\infty} (-1)^{n+1} \frac{\left( \Omega_{0} - \dfrac{\chi_{0}-1}{\tau} \right) \left( 2\Omega_{0} - \dfrac{\chi_{0}-1}{\tau} \right)...\left( n\Omega_{0} - \dfrac{\chi_{0}-1}{\tau} \right) }{R_{0}^{n} n! (4n^{2}-1)} \, h^{2n+1}   \right)
\end{eqnarray}
Inserting the key function (\ref{funkcja_kluczowa_dla_przypadku_f0}) into (\ref{metryka_typu_Nxany}) one gets the metric
\begin{eqnarray}
\label{metryka_przyklad_1}
ds^{2} &=& 2 \phi^{-2} \left\{  \tau^{-1} (d \eta dw - d \phi dt) - \phi (\Omega_{0} t^{-1} - \phi t^{2(\chi_{0}-1)} \dot{U}) dt^{2}  \right.
\\ \nonumber
&& \ \ \ \ \ \ \ \ 
[2\phi^{2} t^{2(\chi_{0}-1)} (h\dot{U} - U) - 2 \Omega_{0} \phi h t^{-1} ]dw dt
\\ \nonumber
&& \ \ \ \ \ \ \ \ 
\left. [\phi^{2}t^{2(\chi_{0}-1)} h (h \dot{U}-2U) + R_{0} \phi t^{-1}] dw^{2} \right\}
\end{eqnarray}
with $U=U(h)$ given by (\ref{wzor_na_pochodna_u}) and $\eta = -h \phi$. The metric (\ref{metryka_przyklad_1}) depends on five parameters: $\chi_{0}, \Omega_{0}, R_{0}, c_{1}$ and $c_{2}$. (Note that the form of metric (\ref{metryka_przyklad_1}) is independent of the last term, $C_{0} t^{1-4\chi_{0}}$ in the key function $W$ given by (\ref{funkcja_kluczowa_dla_przypadku_f0})). 

Differentiating Eq. (\ref{rownanie_przy_f0_2}) over $h$ we get the differential equation for $U$
\begin{equation}
\label{uogolniony_Houser}
(\Omega_{0} h^{2} + R_{0}) \ddot{U} - 2 \frac{\chi_{0}-1}{\tau} h \dot{U} - 2 \left( \Omega_{0} -  \frac{2(\chi_{0}-1)}{\tau} \right) U = 0
\end{equation}
Assuming $\Omega_{0} \ne 0$ and introducing new variable
\begin{equation}
z:= \frac{1}{2} \left( 1-i \sqrt{\frac{\Omega_{0}}{R_{0}}} h \right)
\end{equation}
one brings Eq. (\ref{uogolniony_Houser}) to the hypergeometric equation
\begin{equation}
z(1-z) U_{zz} + \frac{\chi_{0}-1}{\tau \Omega_{0}} (2z-1)  U_{z}  + 2 \left( 1-  \frac{2(\chi_{0}-1)}{\tau \Omega_{0}} \right) U=0
\end{equation}
Finally, we should ask the natural question whether one can obtain the real slices of the $\mathcal{HH}$ space endowed with the metric (\ref{metryka_przyklad_1}). It is quite evident that taking all coordinates, parameters and the function $U$ in (\ref{metryka_przyklad_1}) as the real objects one gets a real metric. However, as can be easily seen this metric is of signature $(++--)$. Thus we immediately find an $[\textrm{N}] \otimes [\textrm{N}]$ real metric of signature $(++--)$ admitting the shearfree, twisting and, by the Raychaudhuri equation, also expanding congruence of null geodesics. Of course, from the physical point of view it would be much more interesting to get a real section of the Lorentzian signature $(+---)$, the more so as Eq. (\ref{rownanie_przy_f0_2}) for $\tau \Omega_{0} = 1-\chi_{0}$ or Eq. (\ref{uogolniony_Houser}) for $\chi_{0}=1$ resemble very much the equation describing the famous Hauser metric \cite{9}. Unfortunately, the $\mathcal{HH}$ space with the metric (\ref{metryka_przyklad_1}) does not admit any Lorentzian slice. This is so because, as can be easily shown (see \cite{64}) the ASD congruence of null strings is \textsl{nonexpanding}. But, from the very beginning we assume that the SD congruence of null strings is \textsl{expanding}. Such a $\mathcal{HH}$ space has no Lorentzian slice.


\setcounter{equation}{0}
\section{The generic case $\dot{f} \ne 0$} 
\label{sekcja_f_niezero}

Here we are going to investigate the $\mathcal{HH}$ system of equations (\ref{uklad_rownan_na_typN_bez_symetrii}) for the generic case when $\dot{f} \ne 0$. Substituting (\ref{rozwiazania_po_uwaglednieniu_K2}) into the system (\ref{uklad_rownan_na_typN_bez_symetrii}) and keeping in mind the relation (\ref{rozwiazanie_na_gamma}) one gets
\begin{subequations}
\label{system_of_equations_for_fne0}
\begin{eqnarray}
\label{rrownanie_1}
&&  \left( P-2h \frac{d Z}{d v} + h^{2} \frac{d^{2} N}{d v^{2}} \right) \ddot{u} + 2  \left( \frac{d Z}{d v} - h \frac{d^{2} N}{d v^{2}} \right) \dot{u}
\\ \nonumber
&& \ \ \ \ \ \ \ \ \ \ \ \ \ 
+ \frac{2 (\chi_{0}-1)}{\tau} (3u - h \dot{u})  = \frac{\varkappa(t)}{2 \tau^{2}} t^{3-2\chi_{0}}
\\ 
\label{rrownanie_2}
&& 2 \left( Z - h \frac{d N}{d v} \right) \ddot{u} + 4 \frac{d N}{d v} \dot{u} + \left( \frac{d Z}{d v}  \right)^{2} - P \frac{d^{2} N}{d v^{2}} + \frac{1}{\tau} \left( P - h \frac{d Z}{d v} \right) 
\\ \nonumber
&&  \ \ \ \ \ \ \ \ \ \ \ \ \ 
+ \frac{2\chi_{0}-1}{\tau} v (h \ddot{u} - 2\dot{u} ) = \frac{\gamma_{0}}{\tau^{2}}  
\\ 
\label{rrownanie_3}
&& \frac{d Z}{d v} \frac{d N}{d v} - Z \frac{d^{2} N}{d v^{2}}
+ \frac{\chi_{0}}{\tau} Z + \frac{1-2\chi_{0}}{2 \tau} v \frac{d Z}{d v} = 0
\end{eqnarray}
\end{subequations}
First, since the left hand side of (\ref{rrownanie_1}) depends on $h$ and the right side on $t$ we can put as before (\ref{rozwiazanie_na_varkappa}). Then as $\dot{f} \ne 0 \ \Longrightarrow \ \dot{v} (h) \ne 0$ one can introduce the new variable $v$ instead of $h$ by
\begin{equation}
v=v(h) \ \Longrightarrow \ h = h(v)
\end{equation}
It is also convenient to introduce the new function
\begin{equation}
\label{definicja_T}
T=T(v) :=  \frac{d N(v)}{dv} +  \frac{1-2\chi_{0}}{2\tau} v
\end{equation}
Then, employing (\ref{zwiazki_U_z_NZP}) and (\ref{definicja_T}) we are able to express $\dot{u}$, $\ddot{u}$ and $\dddot{u}$ in terms of derivatives of the functions $Z$, $P$ and $T$ over $v$. However, the following compatibility conditions must be fulfilled
\begin{equation}
\label{compatibility_conditions}
\frac{d^{3} Z}{dv^{3}} = h \frac{d^{3} T}{dv^{3}} , \ \frac{d^{2} P}{dv^{2}} = h \frac{d^{3} Z}{dv^{3}}
\end{equation}
The crucial point in reduction of the system (\ref{system_of_equations_for_fne0}) is that Eq. (\ref{rrownanie_1}) under (\ref{rozwiazanie_na_varkappa}) follows from Eqs. (\ref{rrownanie_2}) and (\ref{rrownanie_3}). To be more precise, straightforward but tedious calculations show that with (\ref{rrownanie_2}) and (\ref{rrownanie_3}) assumed, the first derivative with respect to $h$ of the left hand side of Eq. (\ref{rrownanie_1}) vanishes. Therefore, if Eqs. (\ref{rrownanie_2}) and (\ref{rrownanie_3}) are satisfied the Eq. (\ref{rrownanie_1}) is trivially satisfied. Gathering all that, inserting (\ref{definicja_T}) and the formulae for $\dot{u}$, $\ddot{u}$ and $\dddot{u}$ into (\ref{rrownanie_2}) and (\ref{rrownanie_3}), and employing also the compatibility conditions (\ref{compatibility_conditions}) one arrives at the following system of ODEs
\begin{subequations}
\label{uklad_rownan_po_podstawieniach}
\begin{eqnarray}
\label{equationss_1}
&& \left( \frac{d^{3} Z}{dv^{3}} \right)^{2} = \frac{d^{3} T}{dv^{3}}   \frac{d^{2} P}{dv^{2}}
\\
\label{equationss_2}
&& 2T \, \frac{d P}{dv}  - P \left( \frac{d T}{dv} + \frac{2\chi_{0}-3}{2 \tau} \right) - 2Z \, \frac{d^{2} Z}{dv^{2}} + \left( \frac{d Z}{dv} \right)^{2} - \frac{\gamma_{0}}{\tau^{2}} = 0
\\ 
\label{equationss_3}
&& T \, \frac{d Z}{dv} - Z \, \frac{d T}{dv}  + \frac{1}{2 \tau} Z=0
\end{eqnarray}
\end{subequations}
Then the $\mathcal{HH}$ metric (\ref{metryka_typu_Nxany}) in coordinates $(v,\phi,w,t)$ has the form
\begin{eqnarray}
ds^{2} &=& 2\phi^{-2} \bigg\{ \frac{1}{\tau}  \left( t^{1-2\chi_{0}}  - \phi \frac{d h}{d v}     \right) dv dw    - \frac{1}{\tau} h \, d\phi dw 
- \frac{1}{\tau} \, d \phi dt  
\\ \nonumber
&& \ \ \ \ \ \ \ \ \ 
-   \left( \phi t^{-1}   \left( \frac{dT}{dv} - \frac{1-2\chi_{0}}{2 \tau}  \right) - \phi^{2} t^{2\chi_{0}-2} \left( h \frac{d^{2} T}{dv^{2}} - \frac{d^{2} Z}{dv^{2}} \right) \right) dt^{2}
\\ \nonumber
&& \ \ \ \ \ \ \ \ \ 
+2  \bigg( t^{-2\chi_{0}} T - \phi h t^{-1} \left( \frac{dT}{dv} - \frac{1-2\chi_{0}}{2 \tau}  \right)
\\ \nonumber
&& \ \ \ \ \ \ \ \ \ \ \ \ \ \ \ \ \ \  \ \ \ 
+ \frac{1}{2} \phi^{2}t^{2\chi_{0}-2} \left( h^{2} \frac{d^{2} T}{dv^{2}} - \frac{dP}{dv} \right) \bigg) dw dt
\\ \nonumber
&& \ \ \ \ \ \ \ \ \ 
+  \bigg(  2t^{-2\chi_{0}} Z + \phi t^{-1} \left( P - 2h \frac{dZ}{dv} \right) 
\\ \nonumber
&& \ \ \ \ \ \ \ \ \ \ \ \ \ \ \ \ \ \  \ \ \ 
+ \phi^{2}t^{2\chi_{0}-2} h \left( h \frac{d^{2} Z}{dv^{2}} - \frac{dP}{dv} \right)    \bigg) dw^{2}  \bigg\}
\end{eqnarray}
where, by (\ref{compatibility_conditions}), $h$ is given as
\begin{equation}
h = \frac{\dfrac{d^{3} Z}{dv^{3}}}{\dfrac{d^{3} T}{dv^{3}}} = \frac{\dfrac{d^{2} P}{dv^{2}}}{\dfrac{d^{3} Z}{dv^{3}}}
\end{equation}
Note that the following restrictions must be imposed on $h_{v}$ and $\ddddot{u}$ (compare with (\ref{zwiazki_U_z_NZP}) and (\ref{conditions_na_niezerowy_twist}))
\begin{eqnarray}
\label{conditions_na_niezerowy_twist_dla_fne0}
&&  h_{v} \ne 0, \ \textrm{and} \ \ddddot{u} \ne 0 \ \Longrightarrow 
\\ \nonumber
&& \frac{d^{3} T}{dv^{3}} \ne 0 , \  \frac{\dfrac{d^{3} Z}{dv^{3}}}{\dfrac{d^{3} T}{dv^{3}}} = \frac{\dfrac{d^{2} P}{dv^{2}}}{\dfrac{d^{3} Z}{dv^{3}}} \ne \textrm{const}
\end{eqnarray}
Now our task is to reduce further the system of equations (\ref{uklad_rownan_po_podstawieniach}). To this end we first observe that the solution of (\ref{equationss_3}) can be written in the form
\begin{equation}
\label{rozwiazanie_na_ZT}
Z(v) = \frac{1}{Q'} , \ T(v) = \frac{1}{2 \tau} \frac{Q}{Q'}
\end{equation}
where $Q=Q(v)$ is a function of $v$ and $Q' \equiv \dfrac{dQ}{dv}$. (Further on the prime "$'$" stands for the derivative $\dfrac{d}{dv}$). Then Eq. (\ref{equationss_2}) can be easily integrated and one get $P(v)$ as 
\begin{eqnarray}
\label{rozwiazanie_na_P}
P(v) &=& \tau Q^{\chi_{0}-1} Q'^{-\frac{1}{2}} \int Q^{-\chi_{0}} Q'^{\frac{3}{2}}
\left( \frac{\gamma_{0}}{\tau^{2}} - 2 Q'^{-2} \{ Q,v \} \right) d v
\\ \nonumber
&=& -2 \tau \frac{Q''}{QQ'^{2}} - 4 \tau \frac{ \chi_{0} }{Q^{2}}  + \tau Q^{\chi_{0}-1} Q'^{-\frac{1}{2}} 
\int  Q^{-\chi_{0}} Q'^{\frac{3}{2}} \left(  \frac{\gamma_{0}}{\tau^{2}}   - 4\chi_{0}(\chi_{0}+1)  Q^{-2}   \right)  d v
\end{eqnarray}
where
\begin{equation}
\{ Q,v \} := \frac{Q'''}{Q'} - \frac{3}{2} \frac{Q''^{2}}{Q'^{2}}
\end{equation}
is the \textsl{Schwarzian derivative}. (We are indebted to Maciej Dunajski for pointing out that some objects arising in the paper are closely related to the Schwarzian derivative). The next step consists in putting (\ref{rozwiazanie_na_ZT}) and (\ref{rozwiazanie_na_P}) into Eq. (\ref{equationss_1}) to obtain the final equation for the function $Q$. The unique inconvenience is here the fact that, as it is seen in (\ref{rozwiazanie_na_P}), the $P(v)$ contains some integral in an explicit form. To avoid this inconvenience one can proceed in other way. Namely, one differentiates Eq. (\ref{equationss_2}) over $v$ and then introduces the second derivative $P''$ from (\ref{equationss_1}). Using the relation
\begin{equation}
T Z'''-ZT''' = \frac{1}{2 \tau Z} (Z'^{2} - 2ZZ'')
\end{equation}
which follows immediately from (\ref{rozwiazanie_na_ZT}) one arrives at the equation
\begin{equation}
\label{alternatywne_rownanie}
\left( T' - \frac{2\chi_{0}-3}{2 \tau} \right) P' - T'' P +  \frac{1}{ \tau Z} \frac{Z'''}{T'''} (Z'^{2} - 2ZZ'') = 0
\end{equation}
This last equation and Eq. (\ref{equationss_2}) constitute the system equivalent to the system consisting of (\ref{equationss_1}) and (\ref{equationss_2}). Differentiating (\ref{alternatywne_rownanie}) with respect to $v$ and employing (\ref{equationss_1}) we extract $P$ as
\begin{equation}
\label{alternatywna_postac_na_P1}
P = \frac{ \left( \dfrac{Z'''}{\tau Z T'''} (Z'^{2}-2ZZ'') \right)'  + \left( T' - \dfrac{2\chi_{0}-3}{2 \tau}  \right) \dfrac{Z'''^{2}}{T'''} }{T'''}
\end{equation}
However, one can also find $P$ directly from (\ref{equationss_2}) and (\ref{alternatywne_rownanie}) considering these equations as the system of two algebraic equations for $P$ and $P'$. The determinant of the system reads
\begin{eqnarray}
\nonumber
\Delta &:=& \det  \left( \begin{array}{cc}
                           T' - \dfrac{2\chi_{0}-3}{2 \tau} & - T''   \\ 
                   2T & -T' - \dfrac{2\chi_{0}-3}{2 \tau}  \end{array} \right) = 
                   2TT'' - T'^{2} + \left(  \frac{2\chi_{0}-3}{2 \tau}  \right)^{2} 
              \\ 
              &=& \frac{Q^{2}}{4 \tau^{2}} ( 2ZZ'' - Z'^{2} ) + \frac{(\chi_{0}-2)(\chi_{0}-1)}{ \tau^{2}} \not\equiv 0
\end{eqnarray}
[We put $\Delta \not\equiv 0$ since otherwise $\Delta' = 2TT'''$ is equal to zero and, consequently, $T'''$ vanishes what contradicts (\ref{conditions_na_niezerowy_twist_dla_fne0})]. 

From (\ref{equationss_2}) and (\ref{alternatywne_rownanie}) we get 
\begin{equation}
\label{alternatywna_postac_na_P2}
P(v) = \frac{\left( T' - \dfrac{2\chi_{0}-3}{2 \tau} \right) \left(  2Z Z'' -Z'^{2} + \dfrac{\gamma_{0}}{\tau^{2}} \right) +  \dfrac{2T}{ \tau Z} \dfrac{Z'''}{T'''} (Z'^{2} - 2ZZ'')   }{\Delta }
\end{equation}
Considering (\ref{alternatywna_postac_na_P2}) and (\ref{alternatywna_postac_na_P1}) one finds the equation
\begin{eqnarray}
\label{HORROR_EQUATION}
&& \left(  \dfrac{2Z'''Q'^{-1}}{\Delta T'''} \{ Q,v \} \right)' 
\\ \nonumber
&& - \frac{1}{2\Delta} [Q Q'^{-2}Q'' + 2 (\chi_{0}-2)] \left[ \dfrac{T'''}{\Delta} \left(  2Q'^{-2} \{ Q,v \} - \dfrac{\gamma_{0}}{\tau^{2}} \right)  + \dfrac{Z'''^{2}}{T'''} \right] = 0
\end{eqnarray}
Straightforward calculations lead to the relations
\begin{eqnarray}
&& Z''' = - Q' ( Q'^{-2} \{ Q,v \} )'
\\ \nonumber
&& T''' = - \frac{1}{2 \tau} \frac{Q'}{Q} \left( \frac{Q^{2}}{Q'^{2}}  \{ Q,v \} \right)'
\\ \nonumber
&& \Delta = -\frac{1}{2 \tau^{2}} \frac{Q^{2}}{Q'^{2}}  \{ Q,v \} + \frac{(\chi_{0}-2)(\chi_{0}-1)}{\tau^{2}}
\\ \nonumber
&& Z'^{2} - 2ZZ''' = 2Q'^{-2} \{ Q,v \}
\end{eqnarray}
Straightforward calculations show that the same equation (\ref{HORROR_EQUATION}) can be found when one gets $P'$ from (\ref{equationss_2}) and (\ref{alternatywne_rownanie}) and then compares this $P'$ with the first derivative of (\ref{alternatywna_postac_na_P2}) over $v$. So the final conclusion is that in the generic case when $\dot{f} \ne 0$ the $\mathcal{HH}$ equation reduces to the fifth-order nonlinear ODE (\ref{HORROR_EQUATION}) for one function $Q=Q(v)$.


\setcounter{equation}{0}
\section{Summary} 

The main results of present paper are:
\begin{enumerate}
\item reduction of the $\mathcal{HH}$ equation for the twisting type $[\textrm{N}] \otimes [\textrm{N}]$ $\mathcal{HH}$ space admitting two (necessary noncommuting) homothetic vectors to one ODE on one holomorphic function (see Eqs. (\ref{rownanie_przy_f0_2}) and (\ref{HORROR_EQUATION}))
\item finding the general solution of Eq. (\ref{rownanie_przy_f0_2}) and then the general twisting type $[\textrm{N}] \otimes [\textrm{N}]$ complex vacuum metric with two homothetic symmetries for the special case $\dot{f}=0$ (see (\ref{wzor_na_pochodna_u}) and (\ref{metryka_przyklad_1})).
\end{enumerate}
Reduction of the vacuum Einstein equations for the twisting type N Lorentzian metric with two homothetic symmetries to some ODE (or ODEs) was studied previously by several authors \cite{13,14,15,16,17,18,19,20,21,22,23,25,28,30,35,36}. In the case when such a reduction is investigated for the twisting type $[\textrm{N}] \otimes [\textrm{N}]$ $\mathcal{HH}$ space the problem is that one must consider the constraints imposed on $C_{ABCD}$ and $C_{\dot{A}\dot{B}\dot{C}\dot{D}}$ separately. This problem simplifies very much when we use the $\mathcal{HH}$ space formalism. It was first shown by J.D. Finley, III \cite{38}. He was able to reduce the $\mathcal{HH}$ equation for the twisting type $[\textrm{N}] \otimes [\textrm{N}]$ $\mathcal{HH}$ space with two homothetic symmetries to two ODEs on two holomorphic functions. By taking appropriate coordinates we find one ODE for one holomorphic function and, moreover, in the special case when $\dot{f}=0$ we are able to find the complete solution of the problem. 

In the generic case $\dot{f} \ne 0$ we get the fifth-order nonlinear ODE which is much more complicated than the previous one and at the moment we are not able to find any solution. However one can note some mysterious role played by the Schwartzian derivative in our equation. This role should certainly be elucidated. (About the role of Schwartzian derivative in general relativity see e.g. \cite{63}).
\newline
\newline
\textbf{Acknowledgments}. We are indebted to Maciej Dunajski for his interest in this work and, in particular, in pointing out the role of Schwartzian derivative in our equations.


\end{document}